\documentclass[11pt]{article}
\usepackage{fullpage}   
\usepackage{geometry}     
\usepackage{graphicx}
\usepackage{amssymb, amsmath}
\usepackage{epstopdf}
\usepackage{subfig}
\title{Fractal-like Distributions over the Rational Numbers in High-throughput Biological and Clinical Data}
\author{Vladimir Trifonov$^{1,2}$, Laura Pasqualucci$^{3,4}$, Riccardo Dalla-Favera$^{3,4,5}$, Raul 
Rabadan$^{1,2}$ \\
\small $^1$Department of Biomedical Informatics,\\
\small $^2$Center for Computational Biology and Bioinformatics, \\
\small $^3$Institute for Cancer Genetics and the Herbert Irving Comprehensive Cancer Center,\\
\small $^4$Department of Pathology and Cell Biology,\\
\small $^5$Department of Genetics and Development,\\
\small Columbia University, New York, NY 10032, USA.}
\date{}                               

\newcommand\Li{\mathrm{Li}}

\newcommand\Q{\mathbb{Q}}
\newcommand\N{\mathbb{N}}
\newcommand\R{\mathbb{R}}

\begin{document}
\maketitle

\begin{abstract}
Recent developments in extracting and processing biological and clinical data are allowing quantitative 
approaches to studying living systems. High-throughput sequencing, expression profiles, proteomics, and
electronic health records are some examples of such technologies. Extracting meaningful information from 
those technologies requires careful analysis of the large volumes of data they produce. In this note,
we present a set of distributions that commonly appear in the analysis of such data. These distributions 
present some interesting features: they are discontinuous in the rational numbers, but continuous in the irrational numbers, and possess a certain self-similar (fractal-like) structure. The first set of examples which we present 
here are drawn from a high-throughput sequencing experiment. Here, the self-similar distributions appear 
as part of the evaluation of the error rate of the sequencing technology and the identification of tumorogenic genomic alterations. The other examples are obtained from risk factor evaluation and analysis of relative disease prevalence and co-mordbidity  as these appear in electronic clinical data. The distributions are also relevant to identification of subclonal populations in tumors and the study of the evolution of infectious diseases, and more precisely the study of quasi-species and intrahost diversity of viral populations. 

\end{abstract}

\begin{figure}[ht]
 \centering
 \includegraphics[width=0.3\textwidth]{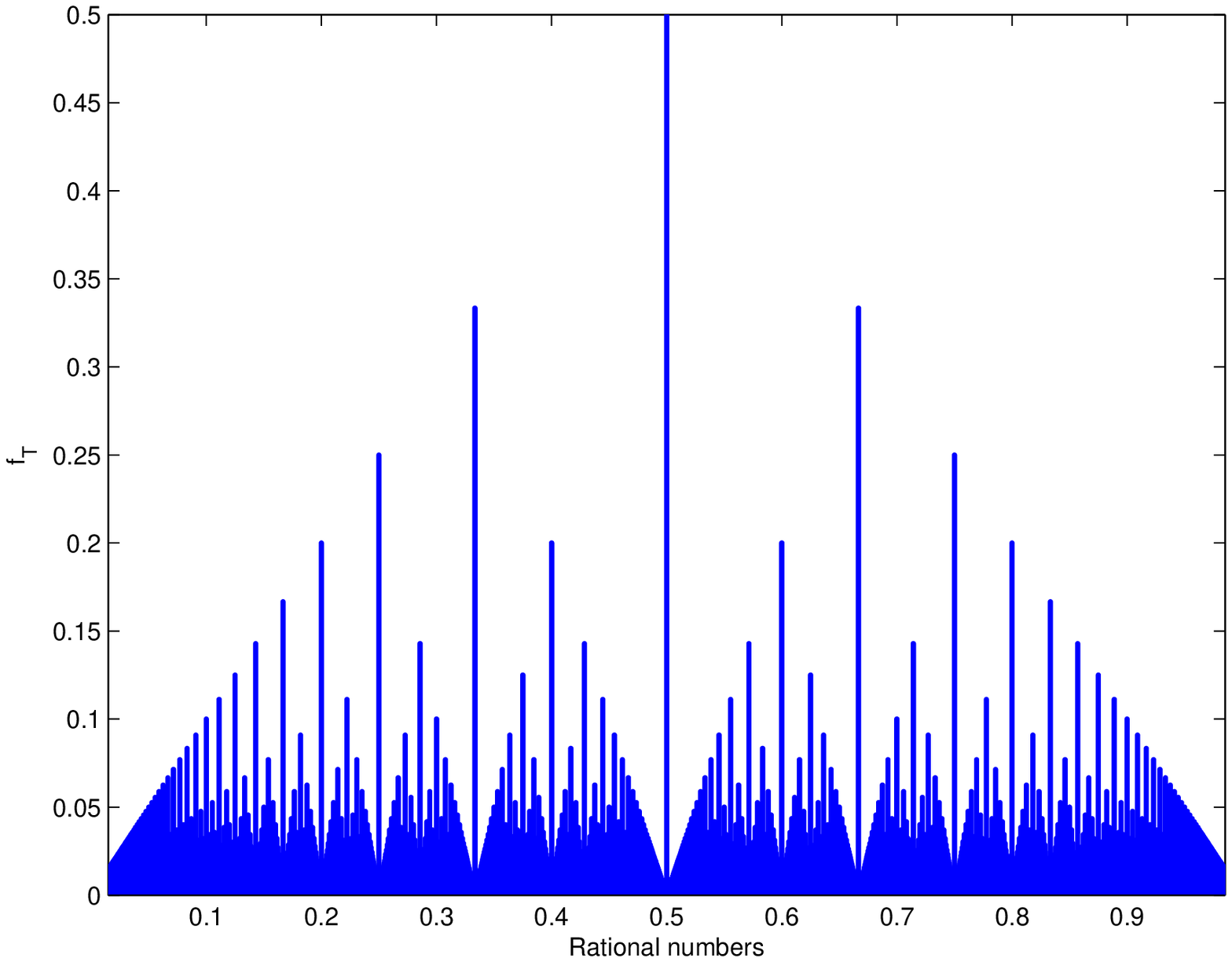}
  \includegraphics[width=0.3\textwidth]{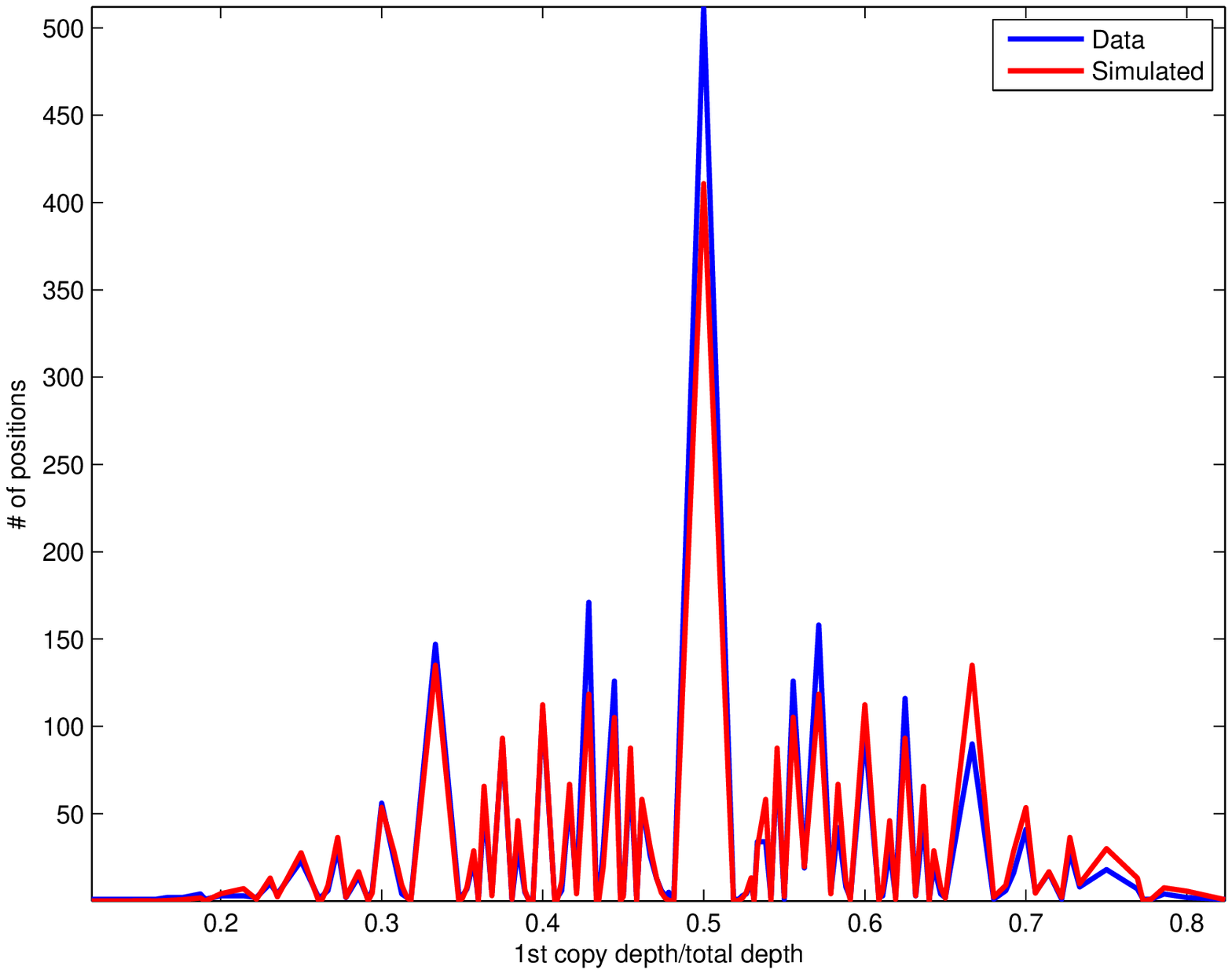}
  \includegraphics[width=0.3\textwidth]{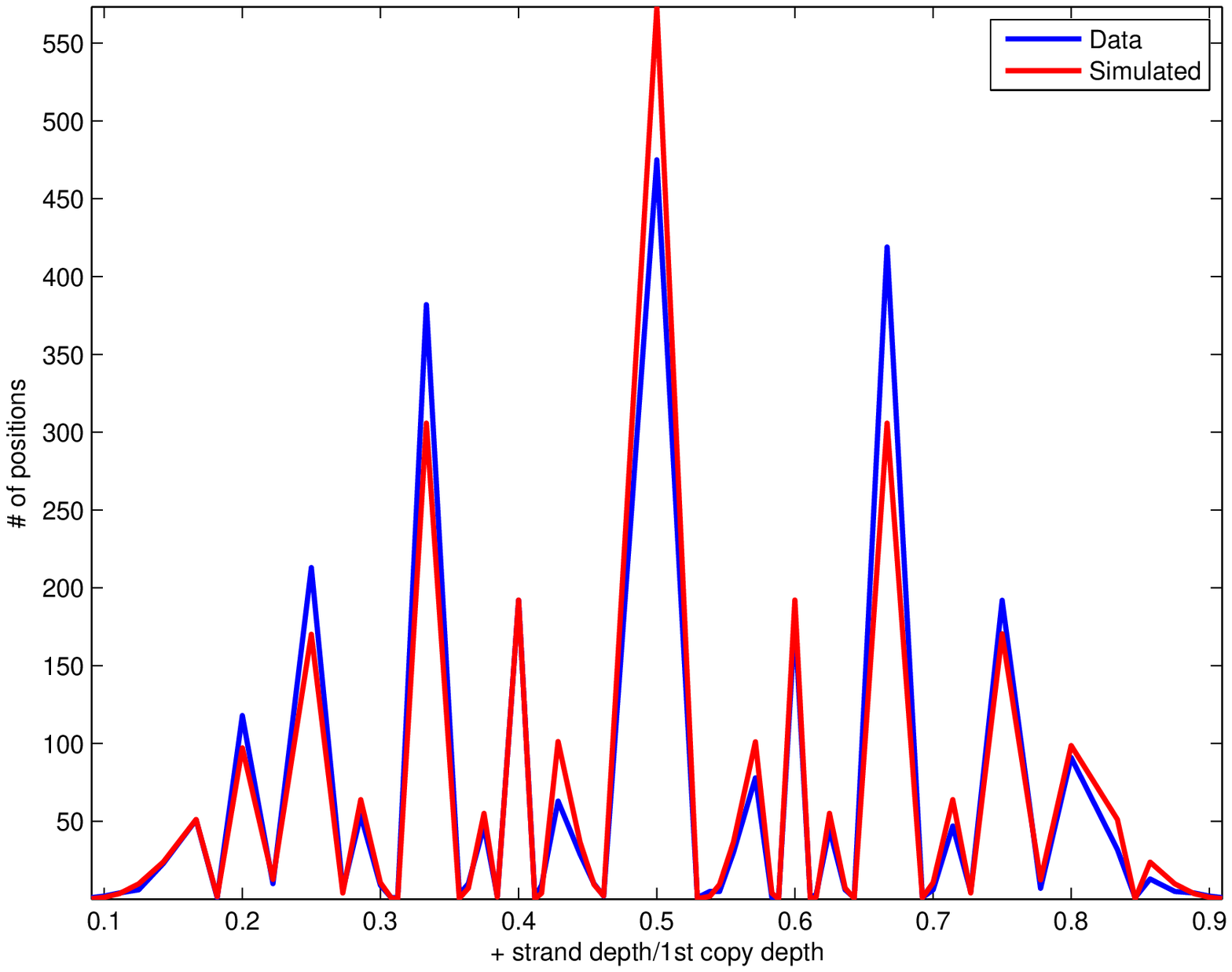}
 \caption{Left: Thomae's function, a self-similar function over the rational numbers in the unit interval. Center/right: Distributions of ratios from heterozygous single nucleotide polymorphisms data from the sequencing of a 
cancer genome.}
   \label{thomae}
\end{figure}

\section{Introduction}
The large volumes of data obtained by recent technological developments, such as high-throughput 
sequencing and expression profiles, are providing novel and complementary ways to studying biological 
systems. In order to extract meaningful, statistically significant information from such data, mathematical methods are being developed, implemented, and tested in various contexts. For example, it is believed  that most tumors are due to somatic mutations that lead to an uncontrolled cell growth. High-throughput sequencing technologies produce hundreds of gigabases of genetic data, providing a way to identify genes responsible for the tumorigenic process by comparing the genome of the tumor and the normal tissue \cite{Bingell09, Mardis10,Ding10, Salk10,Vli10,Pas10}. 

In this note, we point out some interesting properties of the ratios of natural numbers obtained in a
biological/clinical setting. The ratios of interest can be seen as sampled from a distribution over the rational 
numbers in the unit interval. Consider pairs of positive integers, $n$ and $m$, sampled from a distribution 
with probability $f(n,m)$. The ratio $q=n/(n+m)$ of one of these numbers by the sum of the two is a 
rational number in the unit interval. In this way the distribution $f(n,m)$ gives rise to a distribution $g(q)$ 
supported on the rational numbers in the unit interval. A case of particular interest is when the two integers 
are drawn independently from the same distribution $h(n)$. As we are going to see, in this case and for $h
$ being certain common distributions, such as exponential and power-law, it is possible to have a closed-form expression for $g$. We will also see that the resulting distributions over the rational numbers possess certain 
self-similarity properties. Namely, the overall shape of those distributions is similar to Thomae's function 
(Figure \ref{thomae}, left). Although irrelevant to our discussion we would like to point out that, similar 
to Thomae's function, the distributions which we study are rather interesting analytically, because, viewed as functions over the reals, they are continuous on the irrational numbers but not on the rationals. 

We will illustrate the appearance of such distributions in real life data with two examples: 1) a high-throughput 
experiment aimed at identifing genomic variations in cancers and 2) diagnosis information data collected at 
the New York Presbyterian Hospital in several consecutive years. Although the presence of irregular 
shapes and spikes in empirically occuring distributions of ratios of natural numbers was reported before as 
a statistical artifact \cite{John95}, the authors of this previous work failed to acknowledge the interesting 
mathematical structure of the underlying distributions. In this work we propose the study of those naturally 
occurring distributions of rational numbers as an interesting mathematical topic with important clinical and 
biological applications.

\section{First Example: Identifying Genomic Alterations with High-throughput Sequencing}

Our first example comes from a high-throughput sequencing experiment of a diffuse large B-cell lymphoma (DLBCL) sample. DLBCL is the most common B-cell non-Hodgkin lymphoma in adults, accounting for $\approx$40\% of all new lymphoma diagnoses. Tumor DNA was extracted from a nodal tumor of a 63 year old female patient. The coding part the genome (the exome) was enriched using Roche NimbleGen Sequence Capture and the enriched product was sequenced using Roche 454 sequencing.  The data produced from the experiment were $2\cdot 10^6$ reads (sequences of DNA) of average length 250 nucleotides. The reads were aligned to the hg18/NCBI36.1 reference human genome. This resulted in a coverage of about 10x of the human exome and the alignment was used to identify genomic variants distinguishing normal and tumor cells.

\newsavebox{\fgr}
\newlength{\fgrht}

\savebox{\fgr}{
\includegraphics[width=0.3\textwidth]{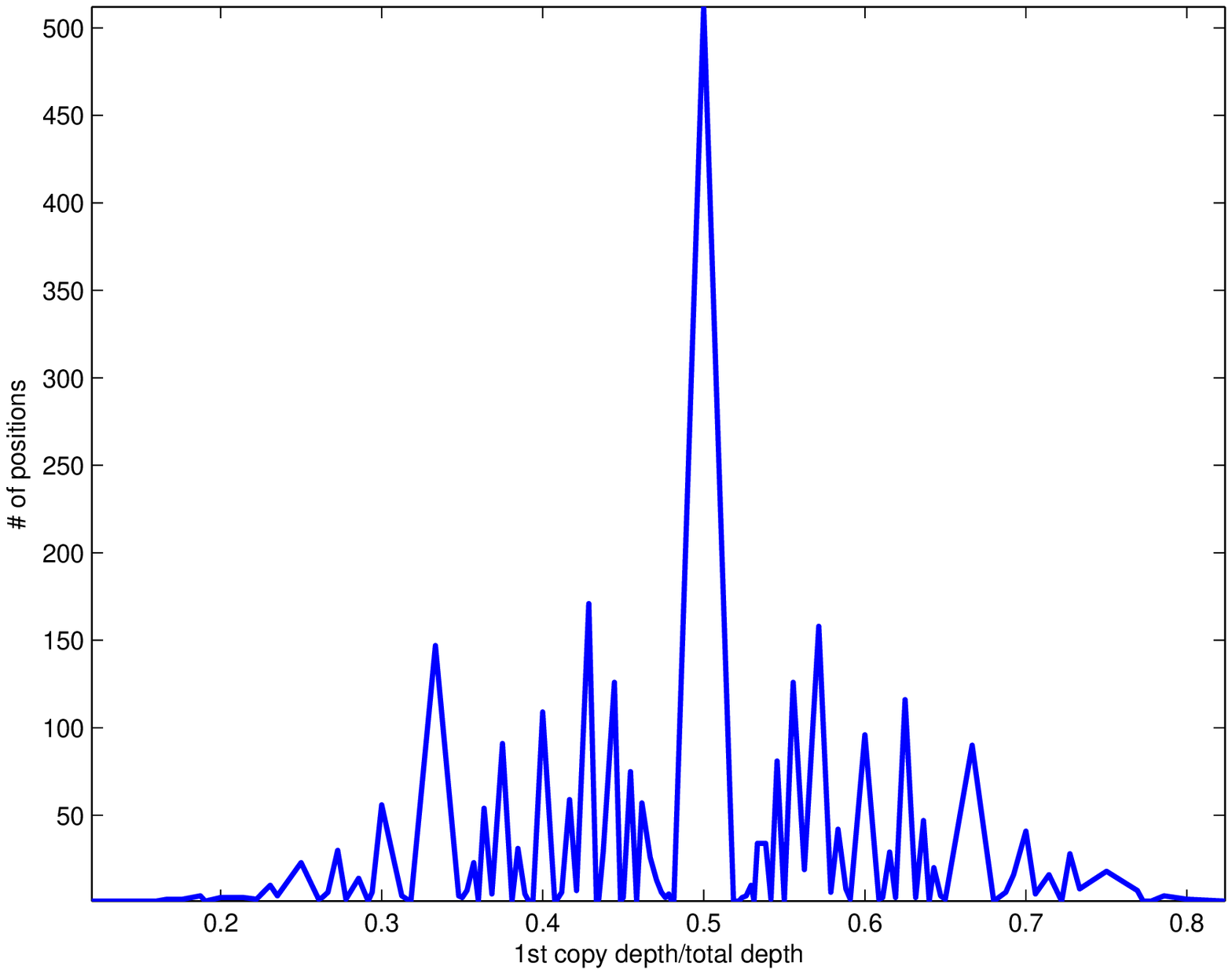}
\includegraphics[width=0.3\textwidth]{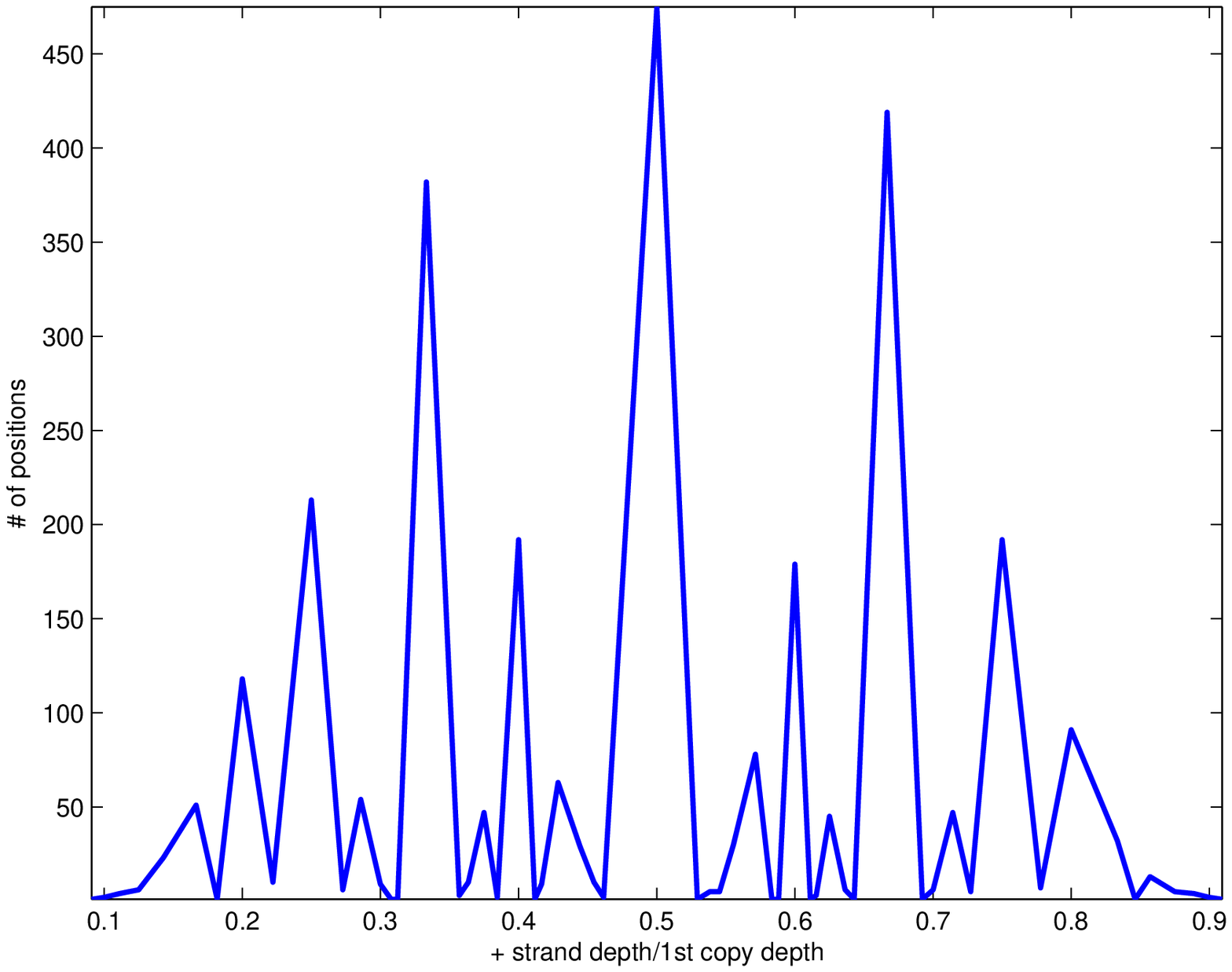}
}
\settoheight{\fgrht}{\usebox{\fgr}}

\begin{figure}[ht]
\centering
\parbox[c][\fgrht]{0.3\textwidth}{\includegraphics[width=0.3\textwidth]{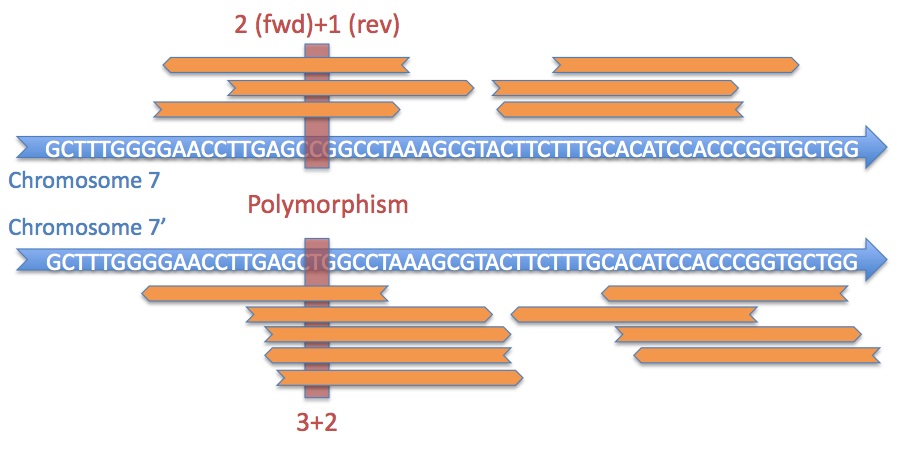}}
\parbox{0.6\textwidth}{\usebox{\fgr}}
\caption{The human genome is diploid with two strands per chromosome. The reads covering a position of 
the genome can originate from each of the four strands. For every position, the ratio between the number 
of reads from one of the strands to the total number of reads from the chromosome and the ratio between 
the number of reads from the chromosome to the total number of reads covering the position are rational 
numbers. The distribution of each of these ratios follows a self-similar distribution.}
\label{reads}
\end{figure}

\begin{figure}[ht]
\centering
\includegraphics[width=0.3\textwidth]{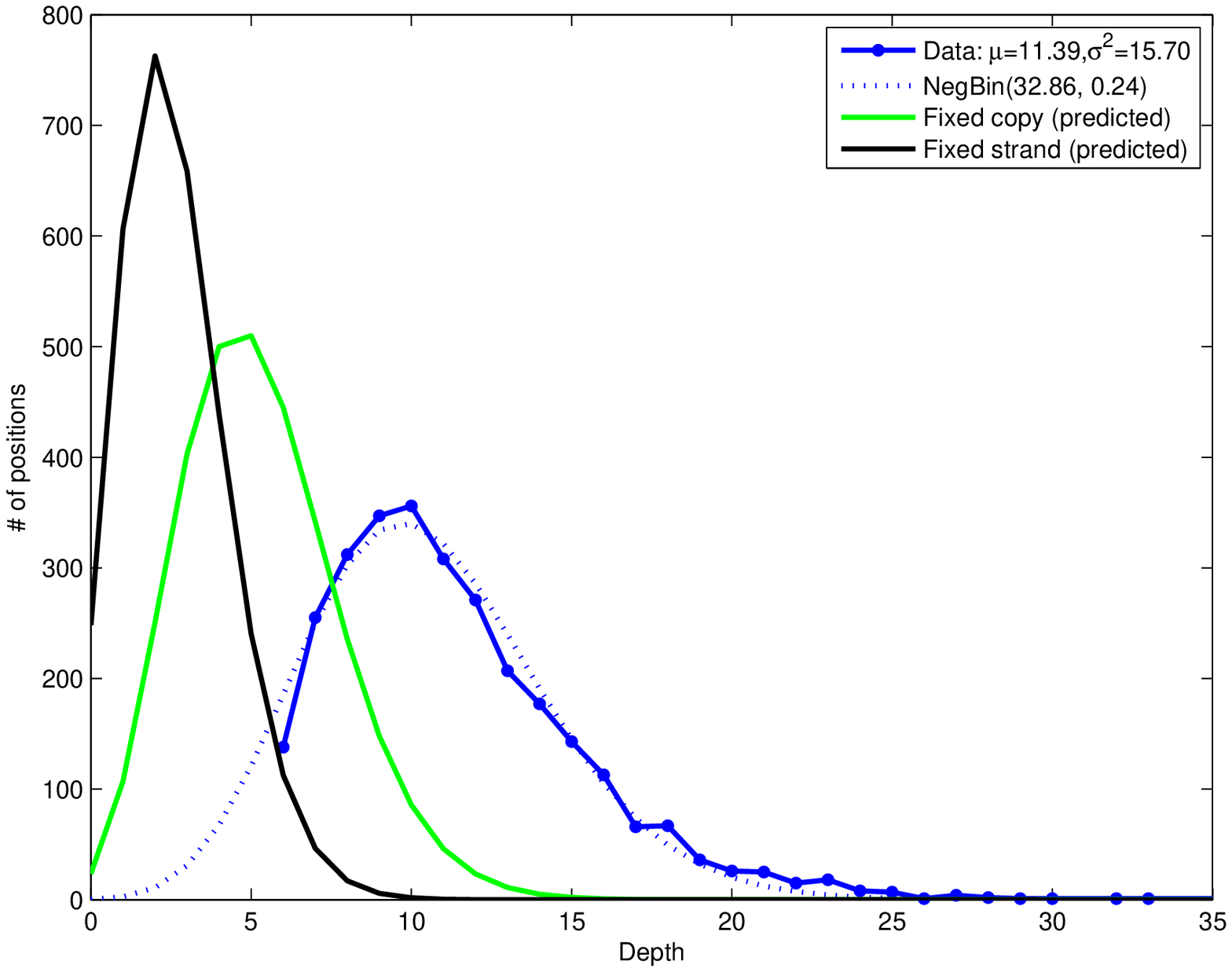}
\includegraphics[width=0.3\textwidth]{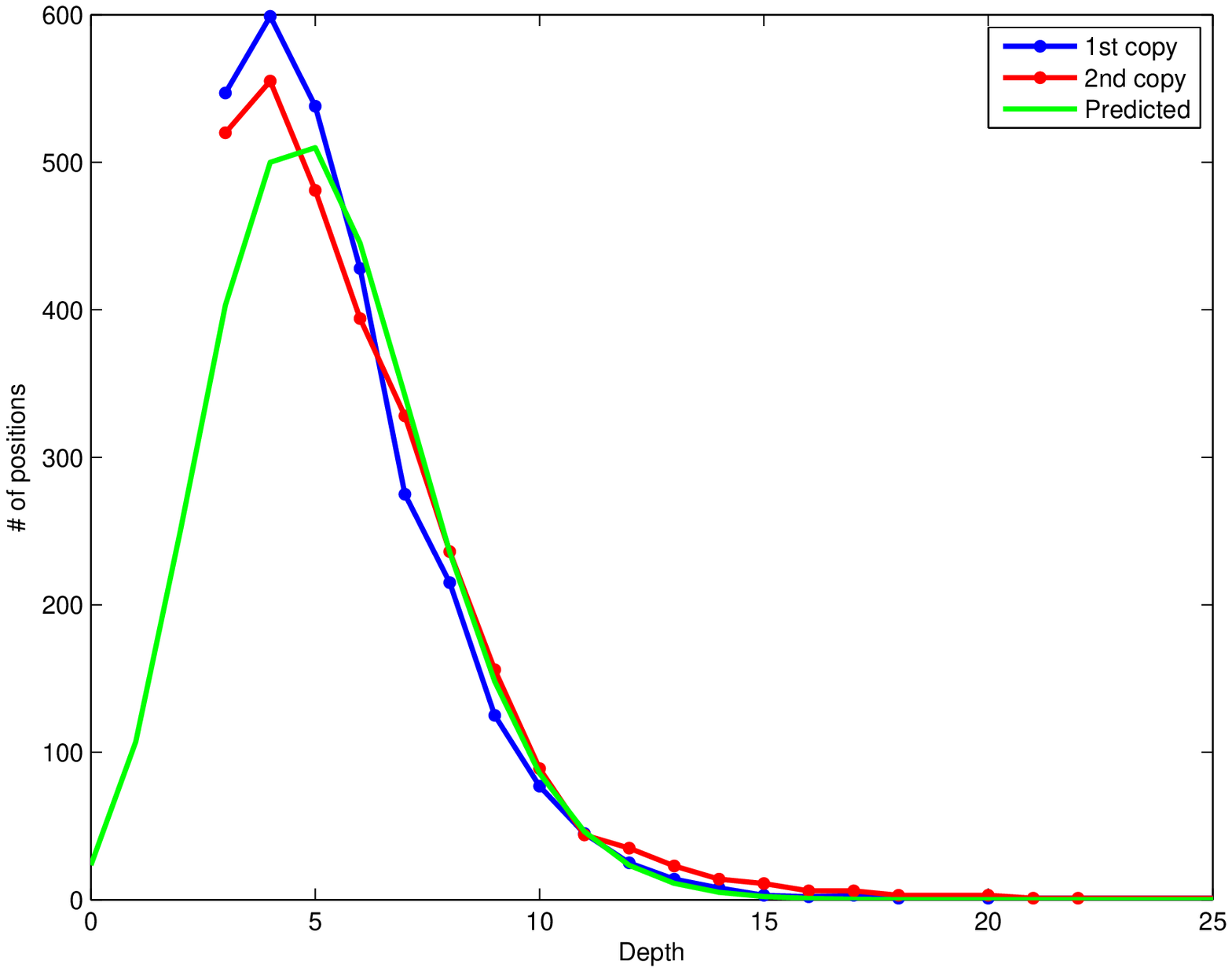} 
\includegraphics[width=0.3\textwidth]{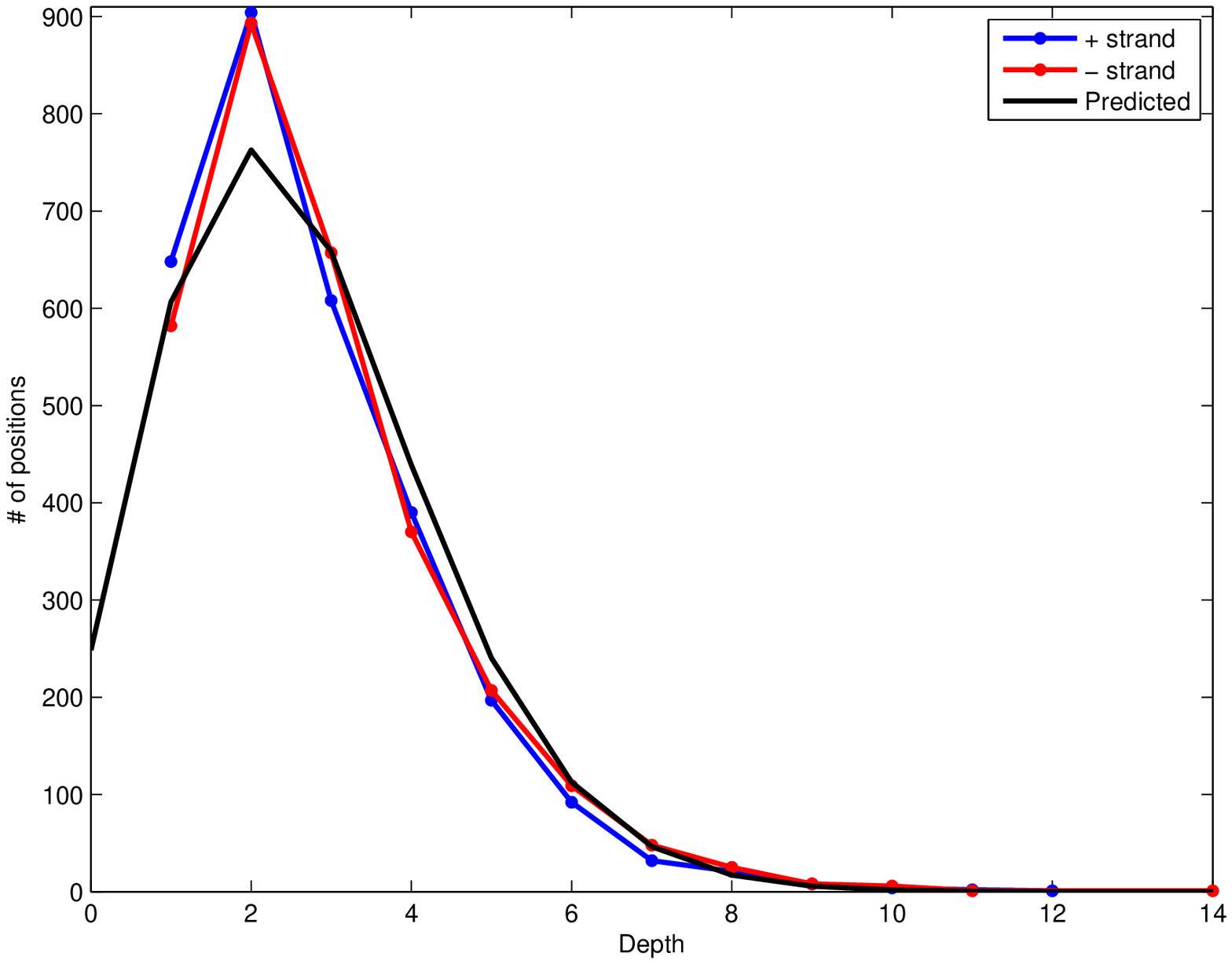} 
\caption{Left: coverage in the cancer sequencing experiment. Center: coverage of the two copies of the 
cancer genome. Right: coverage of the two strands of a fixed copy of the cancer genome.}
 \label{depth}
\end{figure}

Figure \ref{depth} (left, blue) shows the depth (=number of reads covering a particular position) distribution 
(coverage) after alignment of the reads. The figure also shows in red a negative binomial least-square fit of 
the data.  If the reads were obtained from the genome independently and at random, one would expect the 
coverage to follow a Poisson distribution. As it is, even though restricted to a small part of the genome the 
coverage might be Poisson, overall, because of the way the sample was processed before sequencing, the 
means of the Poisson processes in different parts of the genome will vary. The result will be an overdispersion of 
the depth distribution and a better fit by the negative binomial, known to be a mixture of Poisson 
distributions with Gamma-distributed means.

Each of the 46 chromosomes of the human genome has two strands and, with the exception of the sex 
chromosomes X and Y, the human genome is diploid, i.e.\ each chromosome has a homologous copy. 
Since the reference genome is given as entirely haploid, the information about which copy of the genome a 
sample read originates from is not recovered by the alignment. Nonetheless, assuming that a 
read can originate from each copy of the genome with equal probability and given the coverage of the 
reference, one can obtain a theoretical coverage of a fixed copy of the genome. Thus the fraction of positions on a fixed copy of the genome covered with $k$ reads should be
\[
p(k) = \sum_{t=k}^{\infty}q(t)\binom{t}{k} 2^{-t},
\]
where $q(t)$ is the fraction of positions with coverage $t$, as given in Figure \ref{depth} (left, blue). After a 
simple algebraic simplification it can be shown that, if $q$ is $\mathrm{Poiss}(\lambda)$, then $p$ is $
\mathrm{Poiss}(\lambda/2)$. Furthermore, since the negative binomial is a mixture of Poissons with 
Gamma-distributed means, we can obtain that if $q$ is $\mathrm{NegBin}(r,s)$, then $p$ is $
\mathrm{NegBin}(r, (s/2)/(1-s/2))$. Figure \ref{depth} (left, green) shows the theoretical coverage of a fixed 
copy of the human genome obtained from these considerations. Similar reasoning leads us to a predicted coverage of a fixed strand of the human genome shown in Figure \ref{depth} (left, black).

Although the alignment to the reference does not provide exact information about the origin of a read in the 
sample, we can still test the prediction about the coverage of a fixed copy of the cancer genome in the 
following way: take sufficiently many heterozygous positions, i.e.\ positions at which the two copies of the 
genome differ, and then consider the number of reads covering such a position and containing one of the 
variants at that position and the number of reads containing the other variant. Those two depth 
distributions should be close to the predicted distribution of the coverage of a fixed copy of the genome. 
Figure \ref{depth} (center, blue and red) shows the result of these considerations.  Here we took only 
the positions of exonic single nucleotide polymorphisms documented in the NCBI's dbSNP database, 
which are covered sufficiently well in the experiment (total of $\approx$3000 heterozygous positions).  Figure \ref{depth} (center, green) contains the predicted coverage of the two copies of the human genome as obtained earlier. Furthermore, Figure \ref{depth} (right) shows similar plots for a fixed strand of the genome. Since the information about the strand from which a sample read originates is also lost in the sequencing, here we used the orientation of a read when aligned to the reference as a surrogate for its strand. As can be seen, the predictions closely follow the data, confirming our intuition that the reads should come from the four strands of the genome independently.  

Our main observation is concerned with the heterozygous positions we used to obtain the data for Figure 
\ref{depth} (center/right). This time we consider the distribution of the ratios of the number of reads 
covering one of the variants at a particular position in the cancer genome to the total number of reads 
covering this position and the ratio of the number of reads covering one of the strands to the total number 
of reads covering the variant. The resulting distributions of ratios are given in Figure \ref{thomae} (right and 
center, blue).  There are two apparent features of the distributions which drew us to studying them: first, 
their fractal-like self-similar structure, and second, the spikes they contain. We consider the topic of the 
self-similarity of the distributions in Appendix A and quantify it by computing the fractal dimension of 
related functions. From a biological point of view the spikes are interesting because at first sight one might 
decide that they show overrepresentation of certain ratios. For example, for the distribution of variant depth 
over the total depth, the spike at 0.5 is expected, since we are looking at heterozygous positions, but the 
spikes at 0.33 and 0.66 are harder to explain biologically since they would mean the significant presence of 
variants with ploidity other than 2. While such phenomena can occur in cancers because they can 
present genome aberrations known as copy number alterations, the scale at which the phenomenon is 
represented here is unusual. We will see that the spikes are due to the discreteness of the data and could 
actually be explained by a simple stochastic model. Hence regarding the biological conclusions one can 
draw from high-throughput sequencing experiments, the message of our note is that when dealing with biological data the stochastic effects due to the discreteness of the data can be big and attention should be used when 
drawing conclusions lest one confuse such effects with real biological phenomena. A similar conclusion 
was drawn in \cite{John95}. In this note we further study the mathematical properties of the 
resulting distributions.

To formalize the situation we first define the convolution over the rational numbers of two functions defined over the natural numbers. Let 
\[
\Q_u=\Q\cap[0,1]=\{a/(a+b):a\in\N,b\in\N,a+b>0,(a,b)=1\}
\]
be the set of rational numbers in the unit interval. For any two functions $f,g:\N\to\R$ define their convolution 
$c_{f,g}:\Q_u\to
\R$ to be

\[
c_{f,g}(a/(a+b)) = \sum_{m=0}^\infty \sum_{n=0}^\infty f(m)\,g(n)\,\delta\left(\frac{a}{a+b} - \frac{m}{m+n}
\right)=\sum_{t=1}^\infty f(t a) g(t b).
\]

Note that, if $f$ and $g$ are distributions over $\N$, then $c_{f,g}$ is a distribution over $\Q_u$.

In Figure \ref{thomae} (center, red) we have also plotted the convolution $c_{p,p}$ of the negative-binomially distributed predicted coverage $p$ of the two copies of the cancer genome as given in Figure 
\ref{depth} (center, green). In Figure \ref{thomae} (right, red) we have done the same for the coverage of a 
fixed strand. As can be seen, the convolutions closely follow the empirical distributions of ratios. This 
observation is consistent with the null-hypothesis of reads originating from the four strands of the human 
genome independently and covering a particular position on the genome with a negative-binomial 
distribution. No further assumption seems to be necessary to explain the irregular shapes of the ratio 
distributions.

We would like to finish the exposition in this section by noting that the observed structures are not 
particular to the Roche 454 sequencing technology and can be observed in sequencing experiments performed 
with other sequencing platforms, e.g.\ Illumina's Solexa and Life Technologies' SOLiD.

\section{Second Example: Electronic Clinical Data}

The development and implementation of electronic clinical records has made available large amounts of longitudinal clinical data. The primary application of electronic clinical data is to improve the quality of health care provided to the individual patients. Although using this data for uncovering large scale correlations and trends comes secondary to this, the impact such data mining will have on the public health is indisputable. Some specific areas which will be influenced by such analyses are the creation of alert systems for emerging infectious diseases, identification of populations at risk, and measuring the efficacy and efficiency of public health measures. A recent example of this is provided by the 2009 H1N1 influenza pandemic. The first wave of the new influenza strain infected a considerable part of the world population at the end of spring 2009 and the beginning of the summer 2010 \cite{Fras09,Anzic09}. Evaluating the impact of the new pandemic strain on the public health involved analyzing large clinical datasets \cite{Jami09,Cowl10,Khia10}.

\begin{figure}[ht]
\centering
\includegraphics[width=0.3\textwidth]{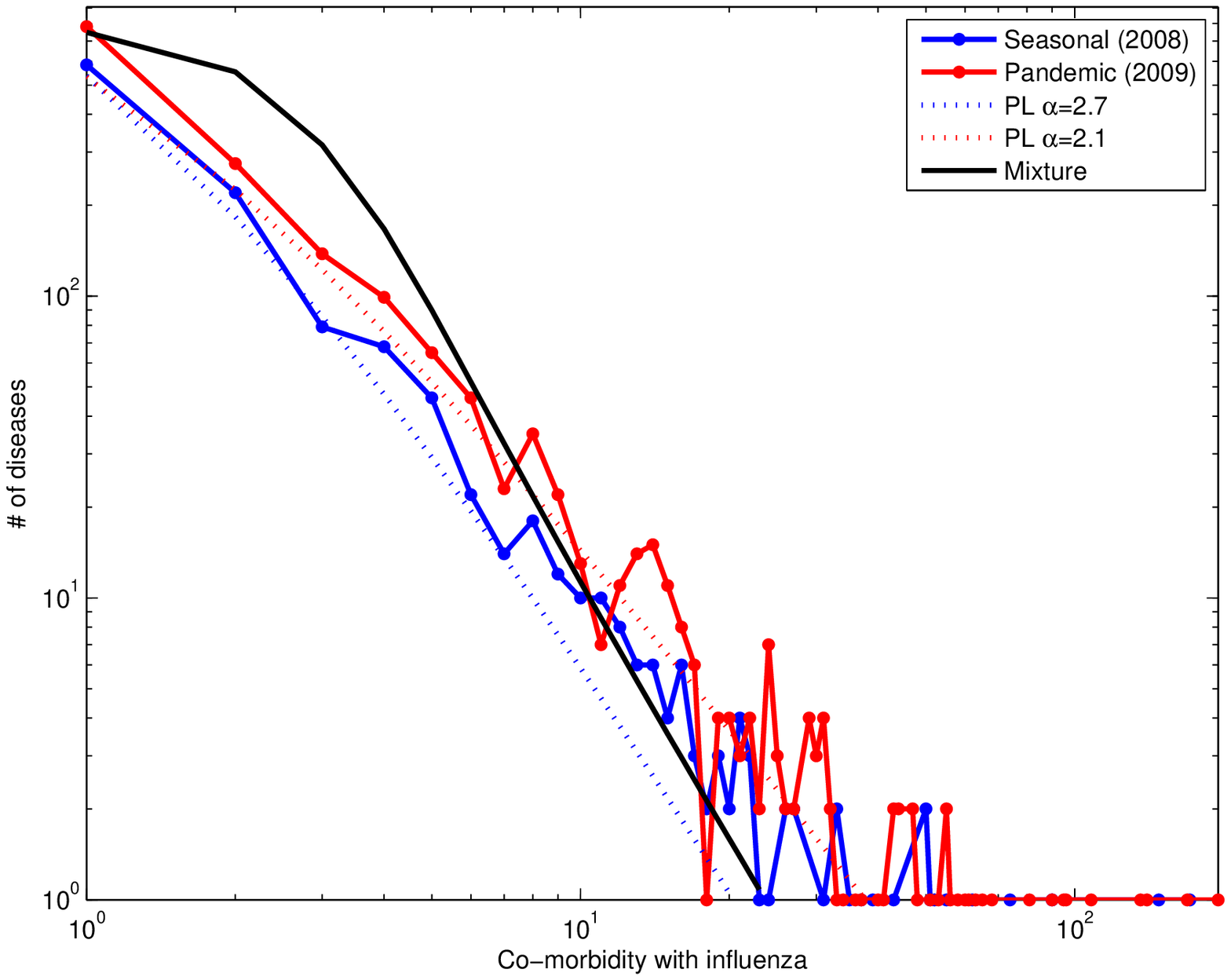}
\includegraphics[width=0.3\textwidth]{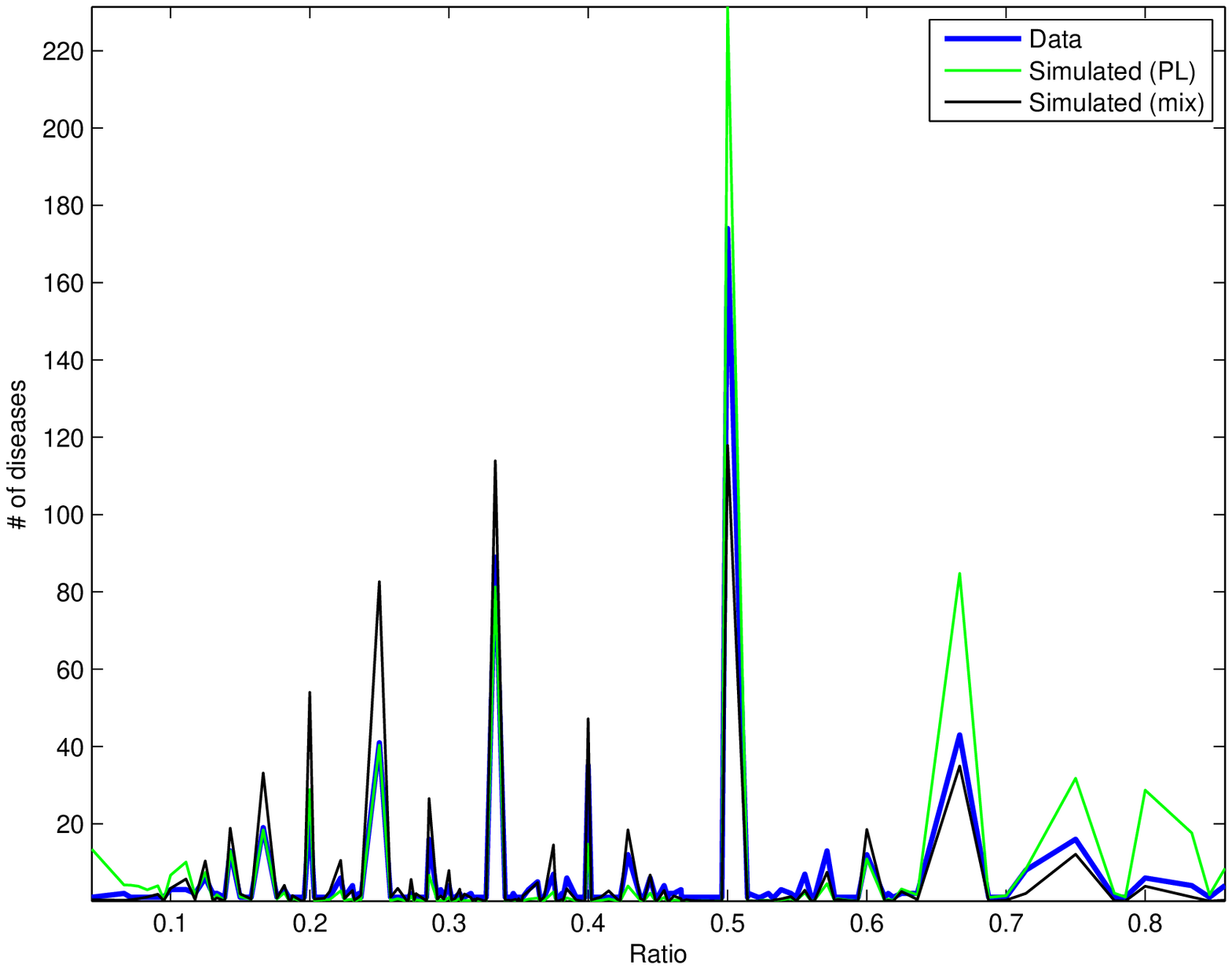} 
\includegraphics[width=0.3\textwidth]{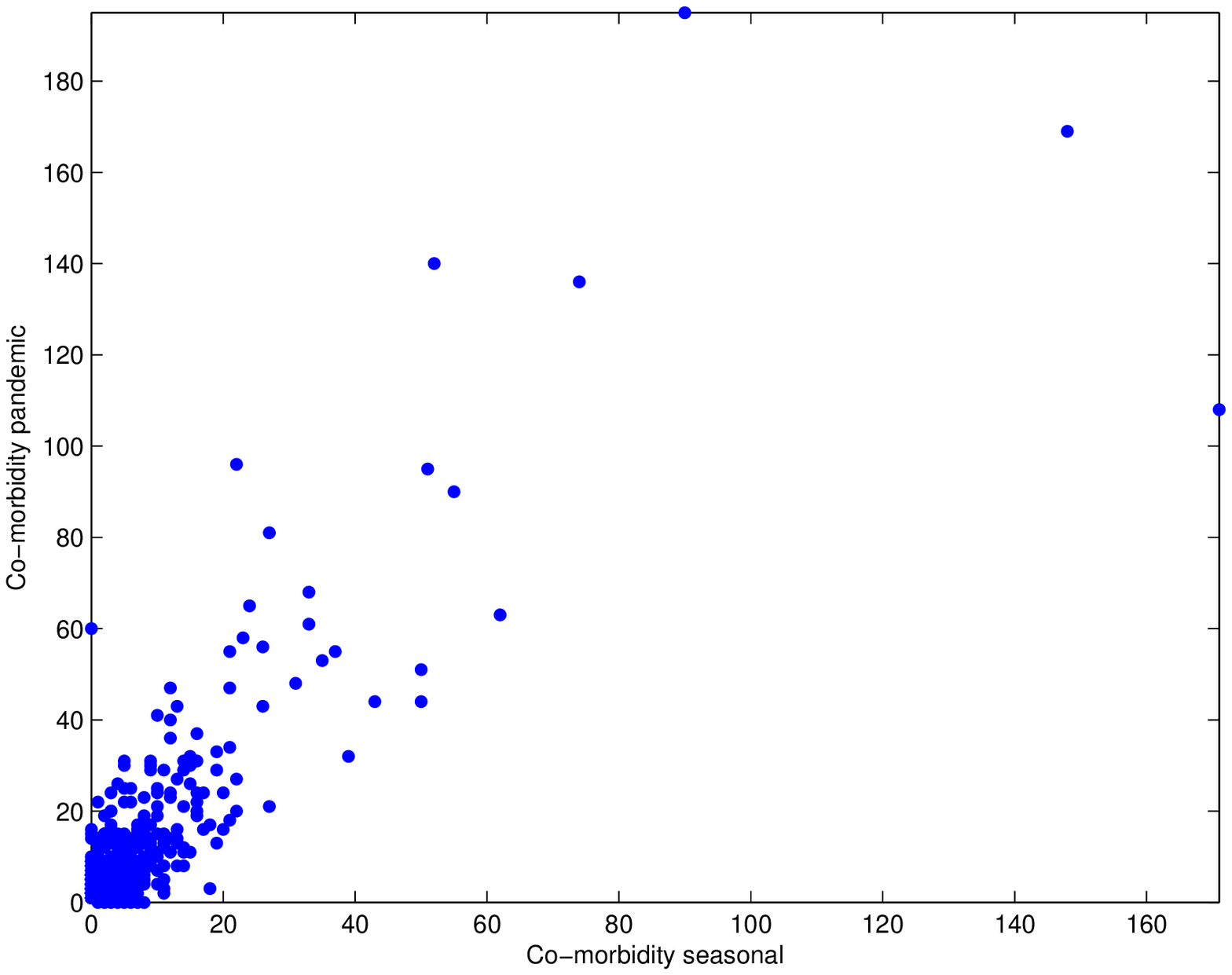} 
\caption{Comparing the co-morbidity of various conditions with the 2009 H1N1 pandemic versus 
seasonal influenza.}
\label{flu}
\end{figure}

The New York Presbyterian Hospital has an electronic repository with the longitudinal clinical records of more than 2 million patients. An example of the large scale analysis enabled by this data is the identification of populations that are at higher risk of morbidity/mortality from the new pandemic influenza virus versus seasonal influenza, for instance, people with asthma, children, pregnant women, etc. \cite{Khia10} The approach we took for this analysis was to compare the number of people with a given condition who were affected by seasonal or pandemic influenza at different time points. Towards this goal, for every two diseases identified by their ICD9 codes, we can obtain from the electronic health records the number of people who have been affected by both diseases. Although this might differ from the established terminology, for the purpose of this note we will call this number the co-morbidity of the two diseases. In this way for a fixed disease we can obtain its co-morbidity with all other possible diseases. If we do this for two diseases, which in our analysis we take to be seasonal and pandemic influenza, we can then compare the sets of co-morbidities and look for conditions enriched with respect to one of the diseases but not the other. Figure \ref{flu} (left) shows the distribution of co-morbidites with seasonal and pandemic influenza. As can be seen these distributions are long-tailed and can be modeled with power-law distributions. The results of the power-law fits are also shown in Figure \ref{flu} (left). 

For a particular health condition, an important measure of the risk of being infected by seasonal versus pandemic influenza for people who have had this condition is the ratio of the number people who have had both that condition and seasonal influenza, i.e.\ the co-morbitity with seasonal flu, to the total number of people who have had the condition, i.e.\ the sum of the co-morbidities with seasonal and pandemic flu. We have plotted the distribution of these ratios in Figure \ref{flu} (center, blue). As can be seen its shape has the self-similar structure of interest to us. From the discussion so far one might be tempted to model this distribution as the convolution of the power-law distributions modeling the two sets of co-morbidities. The result of this attempt is shown in Figure  (center, green). The graph shows that in this case the convolution is not a good model because the empirical ratios are shifted to the left, wheres the convolution is not. In Figure \ref{flu} (right) we have plotted the pairs of co-morbidities for all conditions. The Spearman correlation coefficient for the two sets is 0.83 and linear regression shows that the co-morbidities for pandemic influenza are 1.3 times the corresponding co-morbidities for the seasonal influenza. Hence one might suppose that the discrepancy is due to the fact that the pairs of co-morbidities are not independent -- the convolution defined above assumes that the two distributions are independent.

To avoid this obstacle we reconsidered our model for the distribution of co-morbidities and asked the following question: what is the source of the long-tail of this distribution? Our stipulation is that 1) for a fixed pair of diseases  the co-morbidity is Poisson distributed, if you observe it at different time points; 2) the means of these Poissons vary from pair to pair of diseases; and 3) the distribution of these means is long-tailed. These stipulations are supported by the data in the electronic health records. Furthermore, it is not hard to show that the mixture of Poissons which have a power-law with an exponent $\alpha$ distributed means is a distributions which has a power-law with an exponent $\alpha$ distributed tail (see Appendix C). We use this observation to model the long-tail distribution of the two sets of co-morbidities. In Figure \ref{flu} (left, black) we have plotted the result of a mixture of Poissons with power-law distributed means.

Next we claim that the observed distribution of ratios is a mixture of convolutions of pairs of Poissons where the mixing is with the same power-law distribution used for the distribution of co-morbidities. More precisely, let's say that the co-morbidity of a fixed condition with seasonal influenza is Poisson with mean $\lambda_s$ and its co-morbidity with the pandemic strain is Poisson with mean $\lambda_p$. From our observation on the dependance between the two sets of co-morbidities, we can say that $\lambda_p=\gamma\lambda_s$ for some $\gamma$. Hence the risk ratio of this condition with the two kinds of influenza will be distributed according to the convolution of the two Poissons, which we denote with $R_{\lambda_s}$. Since the mean of $R_{\lambda_s}$ is $\lambda_s/(\lambda_s+\lambda_p)=1/(1+\gamma)$ (see Appendix B), for $\gamma\neq 1$ this mean will be shifted away from 1/2 depending on $\gamma$. Our model of the distribution for pairs of co-morbidites is a power-law mixture of distributions choosing the two co-morbidities independently according to two Poissons, i.e.
\[
f(n,m)=\int_1^\infty g_{\alpha}(\lambda) P_\lambda(n) P_{\gamma\lambda}(m) d\lambda,
\]
where $g_{\alpha}(\lambda)\propto\lambda^{-\alpha}$. Note that although $f(n,m)$ is not a product distribution, i.e.\ its marginals are not independent, it is a mixture of such distributions. Finally, the distribution of risk ratios is given by 
\[
R(a/(a+b)) = \sum_{m=0}^\infty \sum_{n=0}^\infty f(m,n)\,\delta\left(\frac{a}{a+b} - \frac{m}{m+n}\right)=
\int_1^\infty g_{\alpha}(\lambda)R_\lambda(a/(a+b)).
\]
Figure \ref{flu} (center, green) shows the result of these considerations. We observe a good fit between the empirical distribution to the right of 1/2 and the new model and the predicted overall shift of the model to the left. The apparent discrepancy between the empirical and the theoretical model shows that a further investigation of the model is necessary. Since the goal of this note is to give examples of and draw attention to the interesting self-similar distributions appearing in empirical data, rather than to explore one particular example  in detail, we leave the further analysis of the distribution of co-morbidities and the risk ratios derived from them to a future work.

\section{Closed Form for the Convolution}

As a step towards understanding the mathematical properties of functions over the rational numbers in the unit 
interval obtained as the convolution of functions over the natural numbers, we attempted to obtain a closed 
form, i.e.\ in terms of known functions, for some of them. Ideally, given the considerations above, it would 
be interesting to obtain a closed form for the convolution of two negative binomials or two Poissons. 
Unfortunately we were not able to obtain a closed form in those cases. Since a negative binomial is a sum 
of geometric distributions, let us consider the ratio of two geometrically distributed random variables. More 
generally, we will consider a power-law with exponential cut-off of which the exponential distribution (the 
continuous analogue of the geometric distribution) is a special case. Let $g$ be the probability mass 
function of a variable distributed according to a power-law with exponential cut-off with parameters $\alpha,
\beta\geq 0$ such that $\beta>0$ or $\alpha>1$ , i.e.
\[
g(k) = \frac{k^{-\alpha}e^{-\beta k}}{\Li_{\alpha}(e^{-\beta})},
\]
where
\[
\Li_{\alpha}(x)=\sum_{k=1}^\infty k^{-\alpha} x^k.
\]
is the polylogarithm function.  In particular
\[
\Li_{\alpha}(1) = \zeta(\alpha)\text{ and }\Li_0(x^{-1})=\frac{1}{x-1}.
\]
Then
\[
c_{g,g}(a/(a+b))=\sum_{t=1}^\infty g(t a) g(t b)=\frac{(a b)^{-\alpha}\Li_{2\alpha}(e^{-(a+b)\beta})}
{\Li^2_{\alpha}(e^{-
\beta})}.
\]

{\bf Power-law} Take $\beta=0$ and $\alpha>1$. Then
\[
c_{g,g}(a/(a+b))=\frac{\zeta(2\alpha)}{\zeta^2(\alpha)}(ab)^{-\alpha}.
\]

{\bf Exponential} Take $\alpha=0$, $\beta>0$. Then
\[
c_{g,g}(a/(a+b))=\frac{(e^\beta-1)^2}{e^{\beta(a+b)}-1}.
\]

{\bf Uniform} Although this example does not present a distribution appearing naturally in the discussion above, we believe it is fundamental enough to mention here. Furthermore, as discussed in Appendix A, this example is related to Thomae's function, because a certain infinite analogue of it has the same fractal dimension.

For a natural number $L$ let $f_L$ be the probability mass function which is uniform on the set $
\{1,2,\ldots,L\}$, i.e.
\[
f_L(k)=\begin{cases}
1/L,&k\in\{1,2,\ldots,L\}\\
0,&\text{o/w}.
\end{cases}
\]
Then
\[
c_{f_L,f_L}(a/(a+b))=\sum_{t=1}^{\infty} f_L(ta) f_L(tb)=\frac{1}{L^2}\min\left\{
\left\lfloor\frac{L}{a}\right\rfloor, 
\left\lfloor\frac{L}{b}\right\rfloor
\right\}
=\frac{1}{L^2}\left\lfloor \frac{L}{\max(a,b)} \right\rfloor
\]

{\bf Thomae's function} 

\[
f_T(a/(a+b))=1/(a+b).
\]

This function, supported on the rational numbers in the unit interval, is not a distribution. It is a classic 
example of a function which is constant almost everywhere and yet discontinuous on a dense set. It can be 
beautifully interpreted as the view from the corner of Euclid's orchard -- an imaginary orchard which 
contains a tree at every point with integer coordinates. Although it probably is not the convolution of 
functions over the natural numbers, the fact that versions of it appeared in our empirical data was a pleasant 
surprise to us and one of the main motivations for this study. In Appendix A we will show that the graph of 
this function has a fractal dimension 3/2. 

\section{Conclusions}

We have presented a set of self-similar distributions supported on the rational numbers in the unit interval. 
These functions appear pervasively in the analysis of large datasets when models for the distribution of 
ratios of natural numbers are required. The examples presented in this manuscript are drawn from 
next-generation sequencing data obtained as part of a study on the identification of somatic mutations, on one 
hand, and understanding disease co-morbidity as it is reflected in electronic clinical data, on the other. 
One can envisage further applications in clinical and biological settings in which the estimation of a 
frequency or ratio is necessary. Such examples are provided by the detection of subclonal populations in 
tumor samples, e.g.\ as part of a study on resistance to chemotherapy; the study of quasi-species and 
intrahost viral populations, e.g.\ in HIV and influenza; and studies of drug effectiveness, populations at risk 
in a pandemic, and other topics in clinical research approachable through the analysis of 
risk ratios. We hope that our presentation will stimulate further study of the functions presented here 
and provide a bridge between interesting theoretical work and important clinical applications.

\subsection*{Acknowledgments}
We would like to thank Ben Greenbaum, Arnold Levine, Bud Mishra, and Hossein Khiabanian for helpful 
discussions and comments, and Oliver Elliott for help in the preparation of the manuscript. The work of Trifonov and Rabadan is supported by the Northeast Biodefence Center (U54-AI057158) and the National Library of Medicine (1R01LM010140-01). The work of Pascualucci and Dalla-Favera is supported by N.I.H. Grants CA-092625 and CA-37295 (to R.D.-F.) and a Specialized Center of Research grant from the Leukemia and Lymphoma Society (to R.D.-F.) Thanks to Dr.\ Gaidano (University of Novara) for providing the material for the tumor sequencing experiment.

\appendix
\section{Fractal Dimensions}

The distributions we have found in the examples above present  a self-similar fractal structure. 
Distributions are normalizable by definition, i.e. the sum of all segments of figure 1 should be equal to 1. 
We are interested in calculating the fractal dimensions of more general non-normalizable functions. More 
precisely, given a function $f:\Q_u\to\R$, define $G(f)$ be the set of line segments in the plane from $(q, 
0)$ to $(q, f(q))$ for $q\in\Q_u$. We are interested in $\dim G(f)$, the fractal dimension of the set $G(f)$.

If $f$ forms a probability distribution on $\Q_u$, then $\sum_{q\in\Q_u} f(q)=1<\infty$ and so $\dim 
G(f)=1$.

For a given $\alpha\geq 1$, let $f_{\alpha}:\Q_u\to\R$
\[
f_{\alpha}(a/(a+b))=(ab)^{-\alpha}.
\]
From the discussion in Section 4 follows that for $\alpha>1$
\[
\sum_{q\in\Q_u} f_{\alpha}(q)=\frac{\zeta^2(\alpha)}{\zeta(2\alpha)}.
\]
Hence, in this case,  $\dim G(f_{\alpha})=1$. It will be interesting to obtain $\dim G(f_1)$.  The following 
calculations 
from \cite{BCFM98} should help in obtaining this dimension. 

Let $f_T:\Q_u\to\R$ be Thomae's function
\[
f_T(a/(a+b))=(a+b)^{-1}.
\]
We will show that $\dim G(f_T)=3/2$. Since $\max\{a,b\}=\Theta(a+b)$, one can think of Thomae's function 
as the infinite analogue of the convolution of the uniform distribution on $\{1,\ldots,L\}$ extended to $L=\infty
$.

Let $F_n$ be the $n$-th Farey sequence, i.e. $F_n=\{x_0=0<x_2<\cdots<x_{m_n}=1\}$ is the sequence of 
all rational numbers $x_i=a_i/(a_i+b_i)=a_i/c_i\in\Q_u$, such that $a_i$ and $c_i\leq n$, sorted in increasing order. Let $A_n^{(i)}$ be the area of the trapezoid between the $x$-axis and the line segment with points $(x_{i-1},f_T(x_{i-1}))$ and $(x_i,f_T(x_i))$. Then 
\[
2 A_n^{(i)}=(f_T(x_{i-1})+f_T(x_i))(x_i-x_{i-1})=
\frac{c_{i-1} + c_i}{c_{i-1}^2 c_i^2},
\]
where we use that  $x_i-x_{i-1}= 1/c_{i-1} c_i$. 

Let $A_n=\sum_{i=1}^{m_n} A_n^{(i)}$ be the area under the piece-wise linear curve with points from $F_n
$. We will calculate $A_n-A_{n-1}$ for $n\geq 3$. Consider two consecutive members $a_{i-1}/c_{i-1}$ and $a_i/
c_i$ of $F_{n-1}$, which have an element  $y_j=(a_{i-1}+a_i)/(c_{i-1}+c_i)$ of $F_n$ inserted between them. Then $c_{i-1}+c_i=n$ and
\begin{align*}
2(A_{n-1}^{(i)} - A_{n}^{(j)} - A_{n}^{(j+1)})=&
\frac{n}{c_{i-1}^2 c_i^2}-
\frac{c_{i-1} + n}{c_{i-1}^2 n^2}-
\frac{n + c_i}{n^2 c_i^2}=\frac{1}{c_{i-1} c_i n}.
\end{align*}

For every $n>a>0$ if $d=(a,n)$ there exist unique $0<n'< n$ and $0\leq a'<a$ such that $d=(a',n')$, $n'a-a'n=d^2$, $a'<n'$, and $a''=a-a'\leq n-n'=n''$. If $a/n\in\Q_u-\{0,1\}$, then $(a,n)=1$ and we have that $a'/n',a''/n''\in F_{n-1}$ are consecutive and $a/n\in F_n$ is inserted between them. Hence
\begin{align*}
A_n-A_{n-1}=&-\frac{1}{2}\sum_{\substack{a=1 \\ (a,n)=1}}^{n-1} \frac{1}{n' n'' n}
=-\frac{1}{2n}\sum_{\substack{c=1 \\ (c,n)=1}}^{n} \frac{1}{c(n-c)}
=-\frac{1}{n^2}\sum_{\substack{c=1 \\ (c,n)=1}}^{n} \frac{1}{c}
=-\frac{G_n}{n^2},
\end{align*}
where we let
\[
G_n=\sum_{\substack{c=1 \\ (c,n)=1}}^{n} \frac{1}{c}.
\]

Since $A_2=1$ and $\lim_{k\to\infty} A_k=0$ we obtain that
\begin{align*}
A_k =& 1-\sum_{n=2}^k \frac{G_n}{n^2}=\sum_{n=k+1}^\infty \frac{G_n}{n^2}.
\end{align*}

Since
\[
\sum_{b|n} b G_b=\sum_{b|n} \sum_{\substack{c=1 \\ (c,b)=1}}^{b} \frac{b}{c}
=\sum_{d|n} \sum_{\substack{c=1 \\ (c,n)=d}}^{n} \frac{n}{c}=n \sum_{c=1}^n \frac{1}{c}=H_n,
\]
where $H_n$ is the $n$-th harmonic number, from M\"obius inversion follows that
\[
n G_n=\sum_{b|n} \mu(n/b)  b H_b.
\]
We are ready to obtain an asymptotic expression for $A_k$. Namely
\begin{align*}
\sum_{n=k+1}^\infty\frac{1}{n^2}\sum_{b|n}\frac{\mu(n/b)}{n/b} H_b=&
\sum_{c=1}^\infty \frac{\mu(c)}{c}\sum_{\substack{n=k+1\\ c|n}}^\infty \frac{H_{n/c}}{n^2} 
=\sum_{c=1}^\infty  \frac{\mu(c)}{c^3}\sum_{d=\lceil (k+1)/c\rceil}^\infty \frac{H_d}{d^2}\\
\sim&\sum_{c=1}^\infty  \frac{\mu(c)}{c^3}\sum_{d=\lceil (k+1)/c\rceil}^\infty\frac{\ln d}{d^2}
\sim\sum_{c=1}^\infty  \frac{\mu(c)}{c^3}\int_{k/c}^\infty \frac{\ln x}{x^2} dx\\
\sim&\frac{\ln k}{k}.\\
\end{align*}

Let 
\[
\varepsilon_k=\min_i\{x_i-x_{i-1}\}=1/k(k-1),
\]
where the minimum is over the elements of  $F_k$. We need 
\[
N_k=\Theta(A_k/\varepsilon_k^2)=\Theta(k^3 \ln k)
\]
squares of size $\varepsilon_k$ to cover the set  $G(f_T)$. Hence
\[
\dim G(f_T)=\lim_{k\to\infty} \frac{\ln N_k}{\ln \varepsilon_k}=3/2.
\]

Let $F'_k=\{y_0=0<y_2<\cdots<y_{m_k}=1\}$ be the sequence of rational numbers $x=a/(a+b)\in\Q_u$, 
such that $a,b\leq k$, sorted in increasing order. Using similar arguments as above we can show that the 
length $L_{\alpha,k}$ of the curve with points $(y_i,f_\alpha(y_i))$ satisfies
\[
L_{\alpha,k}=\sum_{\substack{a,b=1\\(a,b)=1}}^k (ab)^{-\alpha}\approx\frac{(k^{2(1-\alpha)}-k^{1-\alpha})
\log k}{\zeta(2)(1-\alpha)}.
\]
Let $A_{\alpha,k}$ be the area under the curve with points $(y_i,f_\alpha(y_i))$. Furhermore, let $\delta_k=
\min_i\{y_i-y_{i-1}\}=\Theta(k^{-2})$ and $N_{\alpha,k}$ be the number of squares of size $\delta_k$ 
necessary to cover $G(f_\alpha)$. Since $N_{\alpha,k}=\Theta(A_{\alpha,k}/{\delta_k}^2)=\Omega(\delta_k 
L_{\alpha,k}/{\delta_k}^2)$ we obtain that for $\alpha\in[0,1]$
\begin{align*}
\dim G(f_\alpha)=\lim_{k\to\infty}\frac{\log N_{\alpha,k}}{\log \delta_k}\geq 2-\alpha
\end{align*}
We believe that this lower bound is an equality.

\section{Mean of the Convolution}
In this section we compute the mean of the convolution of distributions on the natural numbers. Consider
two distributions $f,g:\N\to\R$ and their convolution $c_{f,g}:\Q_u\to\R$
\[
c_{f,g}(a/(a+b))=\sum_{t=1}^\infty f(t a) g(t b).
\]
Define $\varphi_{f,g}:\R\to\R$
\[
\varphi_{f,g}(t)=\sum_{n=1}^\infty \sum_{m=0}^\infty f(n) g(m) \frac{n}{m+n}e^{(n+m)t}.
\]
Notice that the mean of the convolution is exactly $\varphi_{f,g}(0)$.

We have that
\[
\varphi_{f,g}'(t)=\left(\sum_{n=0}^\infty n f(n) e^{n t}\right)\left(\sum_{m=0}^\infty g(m) e^{m t}\right)
=\chi'_f(t)\chi_g(t),
\]
where $\chi_f$ and $\chi_g$ are the moment generating functions of $f$ and $g$ correspondingly.

If $f=g$, then solving the differential equation $\varphi'_{f,f}=\chi'_f \chi_f$ we obtain that 
\[
\varphi_{f,f}=(1/2)\chi_f^2
\]
and so the mean of the convolution of a distribution on the natural numbers with itself is 
\[
\varphi_{f,f}(0)=1/2.
\]

It is interesting to obtain a similar result not assuming the equality of $f$ and $g$. For now we present a proof that if $f$ is Poisson with mean $\lambda$ and $g$ is Poisson with mean $\mu$, then the mean of their convolution is $\lambda/(\lambda+\mu)$. The proof uses the fact that the moment generating function of a Poisson with mean $\lambda$ is 
\[
\chi_f(t)=e^{\lambda(e^t-1)}.
\]
Combining this fact with the observations above we obtain
\[
\varphi'_{f,g}(t)= \lambda e^t e^{(\lambda+\mu)(e^t-1)}.
\]
Solving this differential equation gives that
\[
\varphi_{f,g}(t)=\frac{\lambda}{\lambda+\mu}e^{(\lambda+\mu)(e^t-1)}.
\]
Therefore the mean of the convolution of two Poissons is 
\[
\varphi_{f,g}(0)=\frac{\lambda}{\lambda+\mu},
\]
as claimed.

\section{Mixing Poissons}

For $\alpha>1$ let $M_\alpha$ be a mixture of Poissons with power-law with exponential $\alpha$ distributed means, i.e.
\[
M_\alpha(k)=\frac{\alpha-1}{k!}\int_1^\infty x^{k-\alpha} e^{-x}dx.
\]
For $k>>\alpha-1$ we have that
\[
M_\alpha(k)=\frac{(\alpha-1)\Gamma(k-\alpha+1, 1)}{k!}\sim k^{-\alpha}.
\]

\end{document}